\documentclass[9pt,technote]{IEEEtran} 
\usepackage[]{times}
\usepackage{epsfig}
\usepackage{subfigure}
\usepackage{amsmath}
\usepackage{amssymb} 
\usepackage{graphicx}
\newcommand{\bega}{\begin{eqnarray}}
\newcommand{\ega}{\end{eqnarray}}
\newcommand{\bb}{\begin{equation}}
\newcommand{\ee}{\end{equation}}
\newtheorem{defn} {Definition}
\newtheorem{te}{Theorem}

\newtheorem{cor}{Corollary}
\newtheorem{ex}{Example}

\begin{document}

\title{Eliminating Trapping Sets in Low-Density Parity Check Codes by using  Tanner Graph Covers}

\author{ Milo\v{s} Ivkovi\'{c}, Shashi Kiran Chilappagari
and~Bane~Vasi\'{c},~\IEEEmembership{Fellow,~IEEE}}

\maketitle

{ \renewcommand{\thefootnote}{}

 \footnote{M. Ivkovi\'{c} is with the
Department of Mathematics, University of Arizona Tucson, AZ 85721,
USA, e-mail: milos@math.arizona.edu.} \footnote{S. K. Chilappagari
and B. Vasi\'{c} are with Dept. of Electrical and Computer Eng.
Univ. of Arizona. } \footnote{An earlier version of this work
was presented at the 2007 IEEE Intern. Symposium on Information
Theory (ISIT'07).}

 \footnote{This work was supported by grants from INSIC-EHDR and NSF-CCR (Grant
no. 0634969).}

}

\begin{abstract} We discuss  error floor asympotics and present a method  for improving the performance of low-density parity check (LDPC) codes in the high SNR (error floor) region.
The method is based on Tanner graph covers that do not have trapping sets
from the original code.  The advantages of the method are that  it
is universal, as it can be applied to any LDPC code/channel/decoding  algorithm  and  it improves  performance at the
expense of increasing the code length, without losing the code
regularity, without changing the decoding algorithm,  and, under
certain conditions, without lowering the code rate. The proposed
method can be modified to construct convolutional LDPC codes also.
 The method is illustrated by modifying Tanner, MacKay and Margulis
codes to improve performance on the binary symmetric channel (BSC)
under the Gallager B decoding algorithm. Decoding results  on AWGN channel are also presented to illustrate that optimizing codes for one channel/decoding algorithm can lead to performance improvement on other channels.
\end{abstract}

\begin{keywords}
convolutional
LDPC codes, error floor,  Gallager B,  LDPC codes, min-sum decoding algorithm,  Tanner code, trapping sets.
\end{keywords}

\section{Introduction}

\setcounter{footnote}{0}

 The error-floor problem is arguably the most important problem in the theory of low-density parity check
(LDPC) codes and iterative decoding  algorithms.  Roughly, error
floor is an abrupt change in the frame error  rate (FER) performance
of an iterative decoder in the high  signal-to-noise ratio (SNR)
region (see \cite{rich} for more details and \cite{gal}, 
\cite{urb}, \cite{shok} for  general theory of LDPC
codes).

The error floor problem for iterative  decoding on the binary
erasure channel (BEC) is now well understood,  see \cite{di},
\cite{kulkarni} and the references therein.

In the case of the additive white Gaussian noise (AWGN) channel,
MacKay and Postol in \cite{weakMC} pointed out a weakness in the
construction of the Margulis code \cite{marg} which led to high
error floors. Richardson \cite{rich}  presented a  method to estimate error floors of LDPC codes
and presented results on the AWGN channel. He pointed out that the
decoder  performance is  governed by a small number of  likely error events
related to  certain topological structures in the Tanner graph of
the code, called {\it trapping sets} (or {\it stopping sets} on BEC
\cite{di}).\footnote{The necessary definitions will be given in the
next section.} The approach from \cite{rich} was further refined by
Stepanov \textit{et al.} in \cite{misha2}. Zhang {\it et al.}
\cite{lara} presented similar results based on hardware decoder
implementation. Vontobel and Koetter \cite{koetter} established a
theoretical framework for finite length analysis of message passing
iterative decoding based on graph covers. This approach was used by
Smarandache \textit{et al.} in \cite{roxy} to analyze the performance of
LDPC codes from projective geometries \cite{roxy} and  for LDPC convolutional
codes \cite{roxy2}.

An early  account on the most likely
error  events on the binary symmetric channel  (BSC) for codes which Tanner graphs have cycles is given by
Forney \textit{et al.} in \cite{reznik}. Some results on LDPC codes
over the  BSC appear in \cite{roxy}, as well. 

A significant part of the research on error floor analysis  has also
focused on  methods for  lowering the error floor. The two distinct
approaches taken to tackle this problem are (1) modifying the
decoding algorithm and (2) constructing codes avoiding certain
topological structures. Numerous modifications of the sum-product
decoding algorithm  were proposed, see,  for example, \cite{varnica}
and \cite{olgica2}, among others.

Among the methods from the second group, there  have been novel
constructions of codes with high  Tanner graph girth \cite{tan},
\cite{moura}, as it was observed that codes with low girth tend to
have high error floors. While it is true that known trapping sets
have short cycles \cite{misha2}, \cite{shashi}, the example of
projective geometry codes, that have short cycles, but perform well
under (hard decision) iterative decoding, suggests that maximizing
the girth is not the optimal procedure.  As the understanding of the
error floor phenomena and its connection with trapping sets grows,
avoiding the trapping sets directly (rather than short cycles) seems
to be a more efficient way (in terms of code rate and decoding
complexity), to suppress error floors.

Code modification for improving the performance on the binary
erasure channel (BEC) was  studied  by Wang in \cite{perdue}. To the
best of our knowledge,  it is the first paper on code modification
with maximizing  the size of stopping (or trapping) sets as the
objective.  Edge swapping within the code was suggested as a way to
break the stopping sets. The method that we propose is similar.
Roughly speaking, it consists of taking two (or more) copies of the
same code and swapping  edges between the code copies in such a way
that the most dominant trapping sets are broken. It is also similar
to the code constructions that appear in  Smarandache \textit{et
al.} \cite{roxy2}, Thorpe \cite{thorpe}, Divsalar and Jones \cite{jpl} and Kelley,
Sridhara and Rosenthal \cite{deepak}.

The advantages of the method are: (a) it is universal as it can be
applied to any code/channel model/decoding algorithm and  (b) it
improves performance at the expense of increasing the code length
only, without losing the code regularity, without changing the
decoding algorithm,  and, under certain conditions, without lowering
the code rate. If the length of the code  is fixed to $n$, the
method can be applied by taking $t$ copies of a (good) code $C$ of
length $n/t$ and eliminating the most dominant trapping sets of $C$.
The method can be slightly modified to construct convolutional LDPC
codes as well.  The details are given in Section \ref{Section3}.

We apply our method and construct codes based on Margulis
\cite{marg}, Tanner \cite{tan} and MacKay \cite{MC} codes and
present results on the BSC when decoded using  the Gallager B
algorithm \cite{gal}. It is worth noting that the error floor on the
AWGN channel  depends not only on the structure of the code but  also
on implementation nuances of the decoding algorithm, such as
numerical precision of messages \cite{rich}. Since the Gallager B
algorithm operates by passing binary messages along the edges of a
graph, any concern about the numerical precision of messages does
not arise.

The rest of the paper is organized as follows. In Section
\ref{Section2} we introduce   the notion of trapping sets and their
relation to the performance of the code. We explain the proposed
method  in Section \ref{Section3}. We present numerical results in
Section \ref{Section4} and conclude in Section \ref{conclusion}.

\section{Basic Concepts}\label{Section2}

The Tanner graph of an LDPC code, $\cal{G}$, is a bipartite graph
with  two sets of nodes: variable  (bit) nodes and check
(constraint) nodes. The nodes connected to a certain node are referred to as its neighbors. The
degree of a node is the number of its neighbors. The girth $g$ is
the length of the shortest cycle in $\cal{G}$. In this paper,
$\bullet$ represents a variable node, $\square$ represents an even
degree check node and $\blacksquare$ represents an odd degree check
node.

The notion of trapping sets was first introduced in  \cite{weakMC}, but here we
follow the formalism from \cite{olgica2}.

\begin{defn} For a given $m \times n$ matrix $U= (U_{i,j})$ with $1
\leqslant i \leqslant m,$ $1 \leqslant j \leqslant n,$ {\it the
projection} of a set of $h$ columns indexed by $j_1, j_2, \dots,
j_h$ is an $m \times h$ matrix consisting of the elements  $u_{i,j},$
$1 \leqslant i \leqslant m,$ $j=j_1, j_2, \dots, j_h.$
\end{defn}

\begin{defn} Let $H$ be a parity check matrix of an LDPC code. An
$(a,b)$ {\it trapping set}  $\mathfrak{T}$ is a set of $a$ columns
of $H$ with a projection that contains $b>0$ odd weight rows.
\end{defn}

The definition of the trapping set above is purely topological, that
is, a trapping set can be seen as a subgraph of the  Tanner graph.
In other words, an $(a,b)$ trapping set $\cal{T}$ is a subgraph with
$a$ variable nodes and $b$ odd degree checks. The most probable
noise realizations  that lead to decoding failure are related to
trapping sets (\cite{rich}, \cite{misha2}). A measure of noise
realization probability is referred to as {\it pseudo-weight.}
Following the terminology in \cite{misha2}, an {\it instanton} can
be defined as the most likely noise realization that leads  to
decoding failure.

The instantons on the BSC consist of the received bit
configurations with minimal number of erroneous bits that lead to
decoding failure. Following \cite{shashi}, the notion specific to
BSC, analogous to pseudo-weight, can be defined as:

\begin{defn} The minimal number of variable  nodes that have to be initially in error
for the decoder to end up in the trapping set $\mathfrak{T}$ will be
referred to as  {\it the critical number} $k$ for that trapping
set.\end{defn} \textit{Remark:} To ``end up'' in a trapping set
$\mathfrak{T}$ means that, after a  finite  number of iterations,
the decoder will be in error, on at least one variable node from
$\mathfrak{T},$ at every iteration. Note that the variable nodes
that are initially in error do not have to be within the trapping
set.

We illustrate the above concepts with an example.
\begin{figure}[th]
\centering
\subfigure[(5,3) trapping set] 
{   \label{53} \includegraphics[width=0.83in]{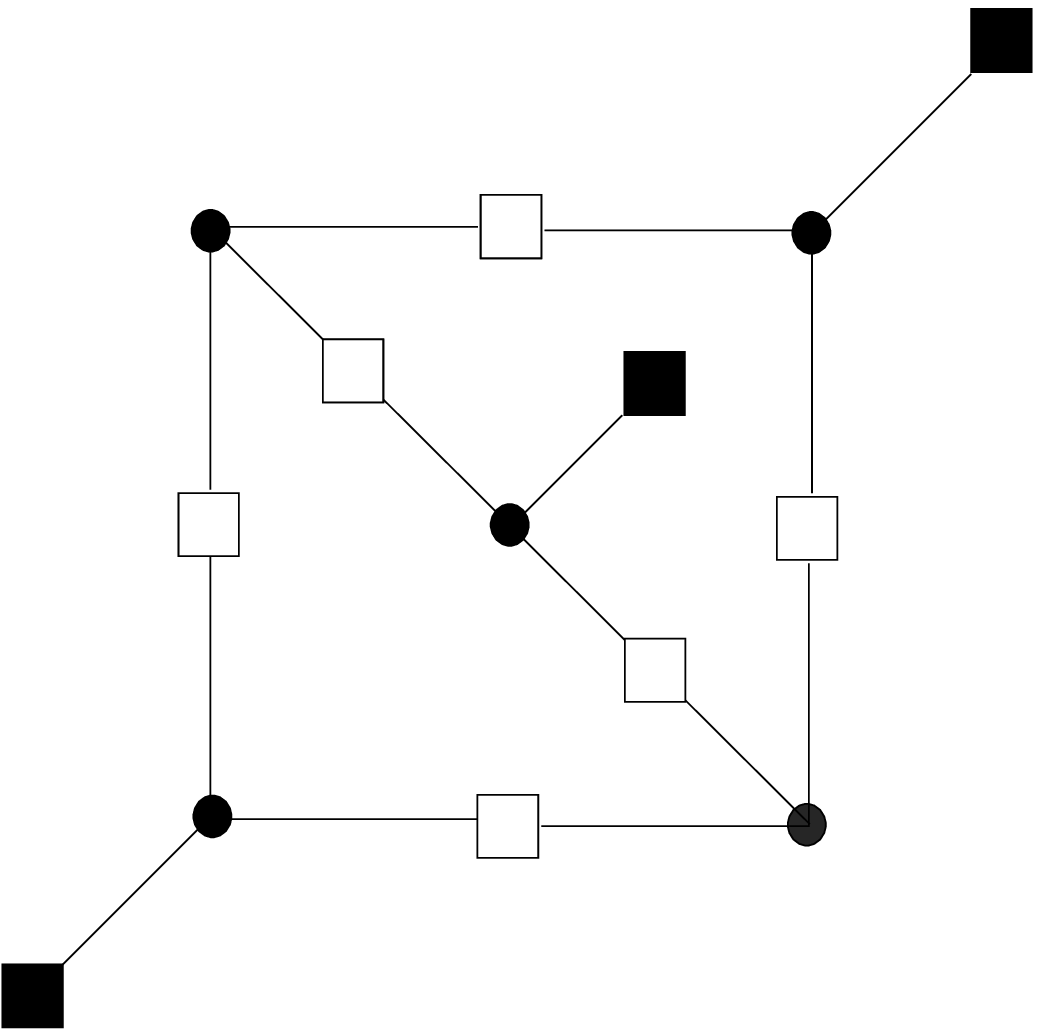} }
\subfigure[(4,4) trapping set]  
{ \label{44} \includegraphics[width=0.84in]{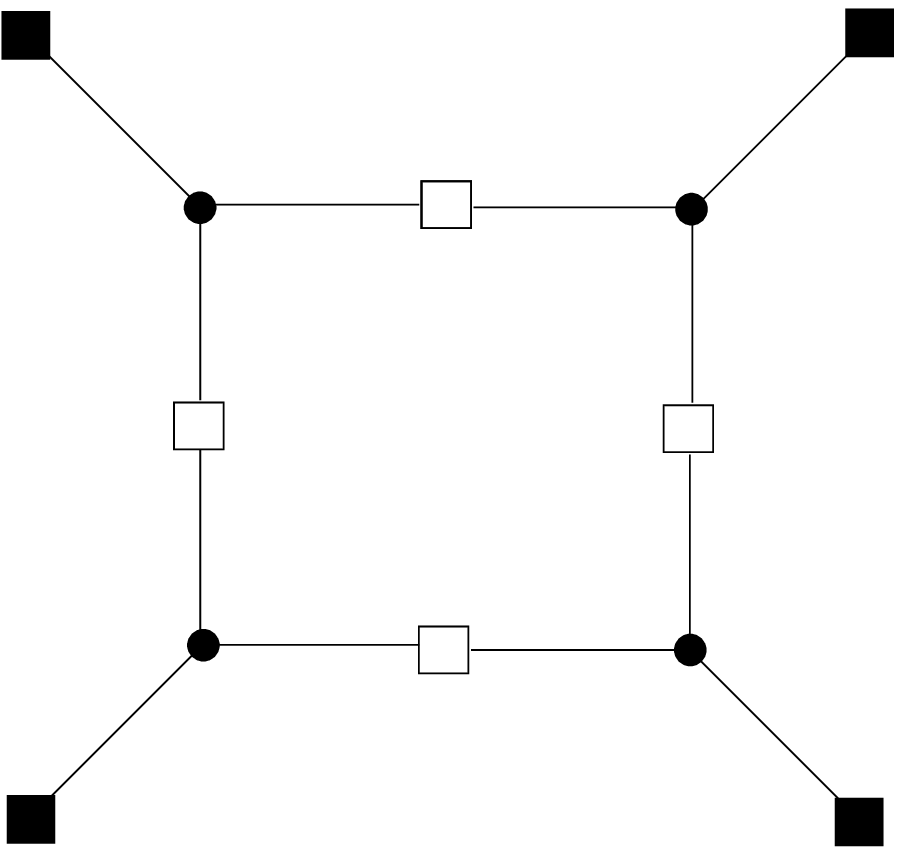} }
\caption{Trapping sets}
\end{figure}
\begin{ex} The $(5,3)$ trapping set in Fig. \ref{53}. appears
(among other codes) in the Tanner (155, 64) code \cite{shashi} (see also the  examples of {\it irreducible closed walks} in the chapter 6.1 of \cite{wiberg}) . This
trapping set has critical number $k=3$ under the Gallager B decoding
algorithm (for the definition of the algorithm see \cite{urb}), meaning that, if three variable nodes, on
the diagonal from bottom left to top right, are initially in error,
the decoder will fail to correct the  errors.

Fig. \ref{44} illustrates a $(4,4)$ trapping set.  This trapping
set, although smaller, has critical number $k=4,$ (all the variable
nodes have to be in error initially for the decoder to fail). So, if
a code has both $(5,3)$ and $(4,4)$ trapping sets, the FER
performance is dominated by the $(5,3)$ trapping set.

At the end of this example, we note that the  $(5,3)$ trapping set
above is an example of an {\it oscillatory trapping set,} i.e, if
three variable nodes on the diagonal are initially in error, after
the first iteration those three nodes will be decoded correctly, but the
remaining two will be in error. In the decoding attempt after the
second iteration those two will be correct, but the initial three
will be in error again, and so on.
\end{ex}
\textit{Remark:} Note that on the BEC the  critical number is just the size of the stopping set, see \cite{perdue}.

We now clarify what ``the most dominant trapping sets'' means and how these effect code performance.

Let $\alpha$ be the transition probability   of the BSC and $c_k$ be
the number of configurations of received bits for which   $k$
channel errors lead to a codeword (frame) error. The frame error
rate (FER) is given by:$$FER(\alpha)=\sum_{k=i}^n
c_k\alpha^k(1-\alpha)^{(n-k)}$$ where $i$ is the minimal number of
channel errors that can lead to a decoding error (size of
instantons) and $n$ is the length of the code.

On a semilog scale the FER is given by the expression
\begin{eqnarray}
\log(FER(\alpha))=\log \big(\sum_{k=i}^n c_k\alpha^k(1-\alpha)^{n-k}\big)\\
=\log(c_i)+i\log(\alpha)+\log((1-\alpha)^{n-i})\\
+\log\left(1+\frac{c_{i+1}}{c_i}\alpha(1-\alpha)^{-1}+\ldots+\frac{c_{n}}{c_i}\alpha^{n-i}(1-\alpha)^{i-n}\right)
\label{other}
\end{eqnarray}
In the limit $\alpha \rightarrow 0$ we note that
$$ \lim_{\alpha \rightarrow 0}
\Big[\log((1-\alpha)^{n-i})\Big]=0 $$
and
 $$ \lim_{\alpha
\rightarrow 0} \Big[\log \Big(1+
\frac{c_{i+1}}{c_i}\alpha(1-\alpha)^{-1}
\ldots+\frac{c_{n}}{c_i}\alpha^{n-i}(1-\alpha)^{i-n}\Big)
\Big]\!\!=0
$$
So, the behavior of the FER curve for small $\alpha$ is dominated by
$$\log(FER(\alpha)) \approx \log(c_i)+i\log(\alpha)$$

The $\log(FER)$ vs $\log(\alpha)$ graph is close to a straight line with slope equal to $i$ -the minimal critical number or cardinality  of the
instantons.

Therefore, if two codes $C_1$ and $C_2$ have instanton sizes $i_1$
and $i_2,$ such that $i_1 < i_2,$ then the code $C_2$ will perform
better than $C_1$ for small enough $\alpha,$ independent of the
number of instantons, just because $ \log(\alpha) \rightarrow
-\infty$ as $\alpha \rightarrow 0.$ Note also that the critical
number of the most dominant trapping sets cannot be   greater than
half the minimum distance. If it is the case, the performance of the
decoder is  dominated by the minimum weight codewords.

\section{The Method for Eliminating Trapping Sets}\label{Section3}

In this section we present a method to construct an LDPC code
$C^{(2)}$ of length $2n$ from a given code $C$ of length $n$ and
discuss a modification of the method that gives a convolutional LDPC
code based on $C$.

\begin{figure}[t]
\centering
\includegraphics[width=1.8in]{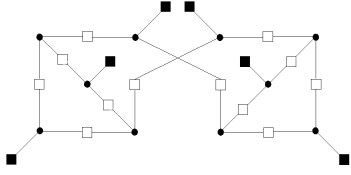}
\caption{Trapping set elimination} \label{killing}
\end{figure}

Let $H$ and $H^{(2)}$ represent the parity check matrices of $C$ and
$C^{(2)}$ respectively. $H^{(2)}$ is initialized to
$$H^{(2)}= \left [\begin{array}{cc} H & 0\\0&H \end{array}
\right].$$ Stated simply, $H^{(2)}$ is formed by taking two copies
of $H$ say $C_1$ and $C_2$. It can be seen that if $H$ has
dimensions $m \times n$, then $H^{(2)}$ has dimensions $2m \times 2n
$. Every edge $e$ in the Tanner graph $\cal{G}$ of $C$ is associated
with a nonzero entry $H_{t,k}$. The operation of changing the value
of $H^{(2)}_{t,k}$ and $H^{(2)}_{m+t,n+k}$ to ``0", and
$H^{(2)}_{m+t,k} \mbox{~and~} H^{(2)}_{m,n+k}$ to  ``1" is termed as
swapping the edge $e$. Fig. \ref{killing} illustrates edge swapping
in two copies of a $(5,3)$ trapping set.  We assume that the most
dominant trapping sets for $C$ are known. The method can be
described in the following steps.

\begin{figure}[h]
\centering
\includegraphics[width=1.8in]{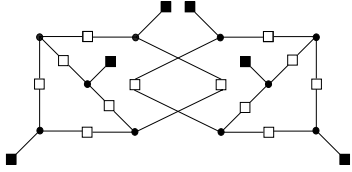}
\caption{Reintroducing trapping set by swapping two edges} \label{notgood}
\end{figure}

{\bf Algorithm:}
\begin{enumerate}
\item Take two copies $C_1$ and $C_2$  of the same code. Since the codes are identical they
have the same trapping sets.  Initialize \textit{SwappedEdges}=$\phi$; \textit{FrozenEdges}=$\phi$;
\item Order the trapping sets by their critical numbers.
\item Choose a trapping set $\mathfrak{T_1}$ in the Tanner graph  of $C_1,$ with minimal critical number. Let $E_{\mathfrak{T_1}}$ denote the set of all edges in $\mathfrak{T_1}$. If ($E_{\mathfrak{T_1}} \cap \mbox{\textit {~SwappedEdges}} \neq \phi $) goto \ref{freezing}. Else goto
\ref{swapping}.
\item \label {swapping} Swap an arbitrarily chosen edge $e \in E_{\mathfrak{T_1}} \setminus \mbox{\textit{~FrozenEdges}}$ (if it exists). Set $\mbox{\textit {SwappedEdges}} = \mbox{\textit {SwappedEdges}~} \cup ~ e$.
\item \label {freezing}``Freeze'' the edges $E_{\mathfrak{T_1}}$ from $\mathfrak{T_1}$ so that they cannot be swapped in the following
steps. Set
$\mbox{\textit{FrozenEdges}}=\mbox{\textit{FrozenEdges}}\cup
E_{\mathfrak{T_1}}$. 
\item Repeat steps 2 to 4 until it is possible to remove the trapping sets of the desired size.
\end{enumerate}
Step \ref{freezing} is needed because swapping additional edges from
the (former) trapping sets  might introduce  trapping sets  with a
same critical number again. Fig. \ref{notgood} illustrates such a
swapping which corresponds to just interchanging the check nodes.

The Tanner graph of  the newly made code is a special double
cover of the original code's Tanner graph,   interested readers are
referred to \cite{koetter}.

\textit{Remark:} There are several approaches  which may improve the
efficiency of the algorithm.  Firstly, instead  of swapping the
edges at random at step 3, edges could be swapped based on the
number of trapping sets they participate in, or by using some other
schedule that would (potentially) lead to the highest number of
trapping sets eliminated. The structure of the code can also be
exploited.  For example, the Margulis $(2640, 1320)$ code
\cite{marg}, has $1320$ $(4,4)$ minimal trapping sets with the
property that each trapping set has one edge that does not
participate in any other minimal trapping set. So, instead of
swapping edges at random, the edges appearing in only one trapping
set can be swapped, and such a procedure is guaranteed to eliminate
all the minimal trapping sets. Also, there is a possibility not to
freeze all the edges from the (former) trapping sets, but only those
that would, if swapped,  introduce the trapping sets with the same
critical number.

Note, however, that any edge swapping schedule can be seen as a
 particular realization of the random edge swapping. For  all the codes that we
considered, all  trapping sets with minimal critical number were
eliminated by the algorithm with random edge swapping.

The following theorem shows how this method  affects the code rate.

\begin{te} If the code $C,$ with parity check matrix $H,$  and rate  $r$ (and length $n$) is
used in the algorithm above, the resulting code $C^{(2)}$ will have rate $r^{(2)}$ (and length $2n$), such that $r^{(2)} \leqslant r.$
\end{te}

\emph{Proof:} Each edge swapping operation in the algorithm can be
seen  as  matrix modification. At the end of the algorithm, code
$C^{(2)}$ is determined by $$H^{(2)}= \left [\begin{array}{cc} H' &
B\\B&H'
\end{array} \right]$$ where $H'$ and $B$ are matrices such that
$H'+B=H,$ and $H'_{t,k}$ (or $B_{t,k}$) can be equal to ``1'' only
if $H_{t,k}=1$.

If the  second block row is added to the first in $H^{(2)}$, and
then the the first block column is added to the second, we end up
with \bega \left [\begin{array}{cc} H' & B\\B&H' \end{array}
\right]\rightarrow \left [\begin{array}{cc} H & H\\B&H' \end{array}
\right] \rightarrow \left [\begin{array}{cc} H & 0\\B&H \end{array}
\right] \label{rank}\ega The last matrix in (\ref{rank}) has rank
which is greater than or equal to twice the rank of $H$. Therefore,
the code $C^{(2)}$ has rate $r^{(2)} \leqslant r$ where $r$ is the
rate of $C$.$\Box$

Note, that $r^{(2)}= r$  if $B=CH+HD,$ for some matrices $C$ and
$D,$ so that $CH$ corresponds  to linear combinations of rows of $H$
and $HD$ corresponds  to linear combinations of columns of $H.$ We
also have  a following corollary.
\begin{cor} If the matrix $H$ has full rank, then $r^{(2)}= r.$
\end{cor}
\emph{Proof:} This follows from the fact that if $H$ has full rank,
then the last matrix in (\ref{rank}) has full rank also. $\Box$

At the end of this section,  we briefly discuss the minimal distance of
the modified code.
\begin{te} If the code $C$ has minimal distance $d_{min},$ the  modified  code $C^{(2)},$
will have the minimal distance $d^{(2)}_{min},$ such that, $2d_{min}
\geq  d^{(2)}_{min} \geq d_{min}.$
\end{te}
\emph{Proof:} We first prove that $d^{(2)}_{min} \geq d_{min}.$
Suppose that the minimal weight codeword of $C^{(2)}$ is $c^{(2)},$
where $c^{(2)}$ is a column vector consisting of two
 vectors $c_1$ and $c_2$ of length $n$. Then $H^{(2)}c^{(2)}=0$ is
equivalent to \bega \left [\begin{array}{cc} H' & B\\B&H'
\end{array} \right]\left [\begin{array}{c} c_1\\c_2
\end{array}\right]= \left[\begin{array}{c} H'c_1+Bc_2\\Bc_1+H'c_2
\end{array}\right] =0 \label{aha}\ega

Note that $c_1+c_2=c$ is a column vector of length $n,$ with Hamming
weight $w_h(c) \leq w_h\left(c^{(2)}\right),$ where
$w_h\left(c^{(2)}\right)$ is the Hamming weight of the $c^{(2)}.$
Now: \bega Hc\!=\!(H'\!+\!B)(c_1\!+\!c_2)= H'c_1\!+\!Bc_1\!+\!
H'c_2\!+\!Bc_2\!=\!0 \label{last}\ega because the last expression in
Eq. (\ref{last}) is equal to the sum of entries of the last column
vector in Eq. \ref{aha}. So, $c$ is a codeword of $C.$

If $c \neq 0$,  from  $w_h(c) \leq w_h\left(c^{(2)}\right)$ it follows that
$d^{(2)}_{min} \geq d_{min}$. If $c=0$ then $c_1=c_2,$ and from Eq. (\ref{aha})  follows that $Hc_1=0,$ so $c_1$ is a codeword of
 $C$ and again $d^{(2)}_{min} \geq d_{min}$.

The proof that  $2d_{min} \geq  d^{(2)}_{min}$ is similar.  If we assume
 that $c_1$ is a minimal weight codeword of $C,$ we have:
\bega  \left [\begin{array}{cc} H' & B\\B&H'
\end{array} \right]\left [\begin{array}{c} c_1\\c_1
\end{array}\right]=0\ega
so $2d_{min} \geq  d^{(2)}_{min}.$

We finish this proof by mentioning  that it is not difficult to
construct examples where $2d_{min} = d^{(2)}_{min}$ or
$d^{(2)}_{min} = d_{min},$ so  the statement of the theorem is
``sharp". $\Box$

We described the algorithm in its basic form.  $H^{(2)}$ can be
initialized by interleaving the copies $C_1$ and $C_2$ in an
arbitrary order, but we choose concatenation to keep the notation
simple. The method, as well as all the  proofs, will hold for any
interleaving. It is also possible to consider more than two copies
of the code to further eliminate trapping sets with higher critical
number.

{ \renewcommand{\arraystretch}{0.7} 

 The splitting of parity check matrix $H$ into $H'$ and $B$ can be seen as a way to construct convolutional LDPC
 codes, that is, as a way to {\it unwrap} the original LDPC code
 $C$. For details on unwrapping see \cite{roxyISIT} and the
 references therein. The (infinite) parity check matrix can be
can be constructed as: \bega  H_{conv}=\left
[\begin{array}{cccc} H'& & &\\B&H'& &\\&B& H'&\\[-5pt]& &B& \ddots \\[-5pt]& &
&\ddots  \end{array} \right]\ega

}

Note that by construction the resulting convolutional code has
pseudo-codewords with higher pseudo-weights than original LDPC code.
In this light, Theorem 2 can be seen as a generalization of Lemma
2.4 from \cite{roxy2}. We refer readers interested in convolutional
LDPC codes to that paper.

\section{Numerical Results}\label{Section4}

In this section we illustrate   the proposed method  by modifying
the Margulis \cite{marg},  Tanner \cite{tan} and MacKay \cite{MC} codes
to eliminate trapping sets  under the Gallager B decoding algorithm.
We use the trapping sets reported in \cite{shashi}.

\begin{figure}[h]
\centering
\includegraphics[width=1.75in]{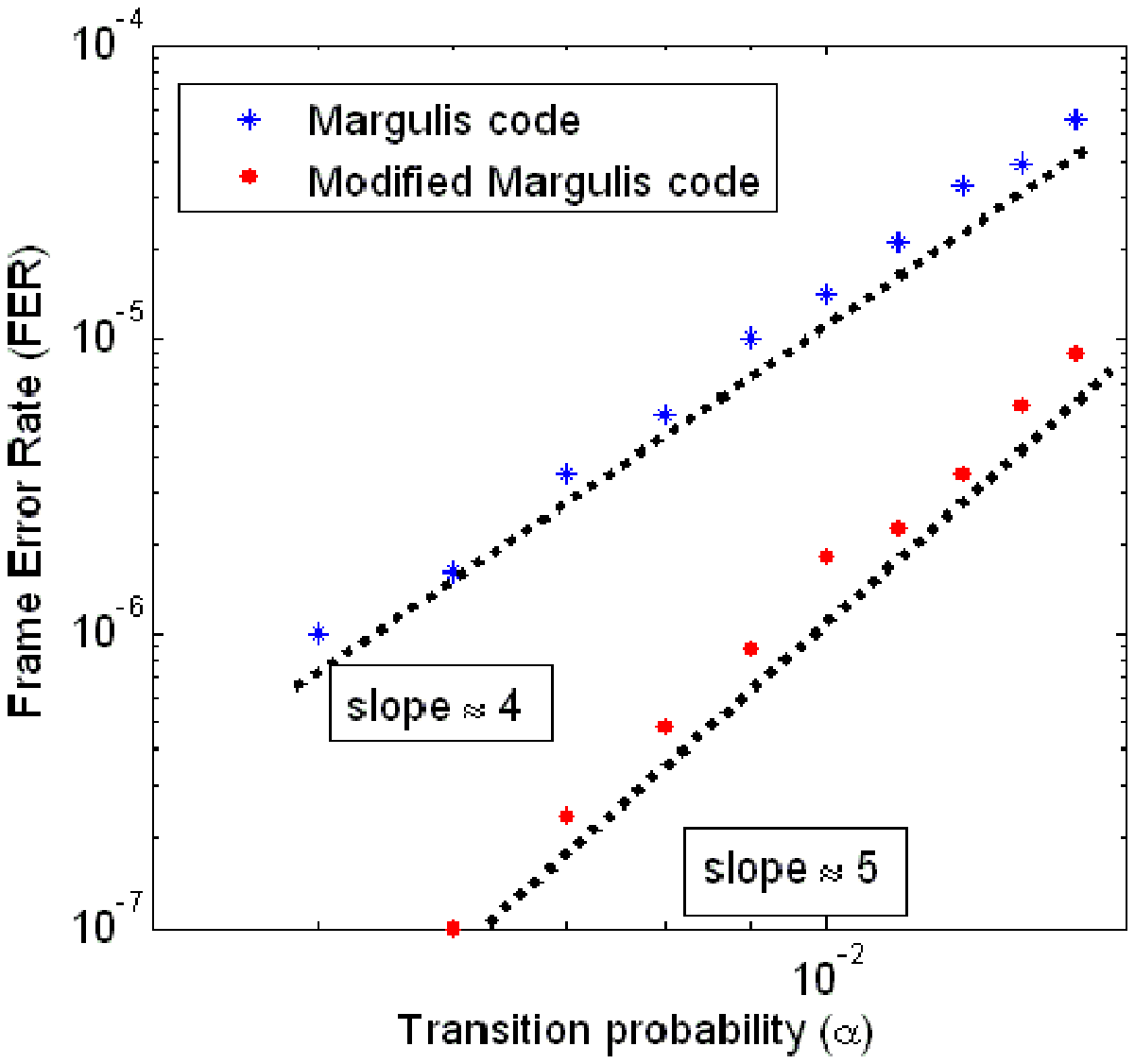}
\caption{Margulis code performance} \label{margulis}
\end{figure}

\begin{ex} {\it (Margulis $(2640, 1320)$ code)} The parity
check of this matrix has full rank, so the modified code is an $(5280,
2640)$ code, and has the same rate as the original code, i.e., $r^{(2)}=r=0.5.$

This code has $1320$  $(4,4)$ trapping sets with critical number $4$
as the most dominant ones. The  modified $(5280, 2640)$ code has no
$(4,4)$ trapping sets and the performance is governed by $(5,5)$
trapping sets (ten cycles), that have critical number $k=5$,
Fig. \ref{margulis}.\end{ex}

\begin{ex} {\it (Tanner (155, 64) code)} This code has $(5,3)$ trapping sets (Fig. \ref{53}) with critical number $i=3$ as
the most dominant ones. There are 155 such trapping sets
\cite{shashi}, \cite{tan}.  In this case we used a version of the method in
which it is possible to swap edges from the (former) trapping sets,
if no trapping set of the same or smaller critical number is
introduced. The result was a (310, 126) code for which the minimal
trapping sets are type (4,4) (eight cycles) with critical number
$k=4$ (see Fig \ref{44}). This was confirmed by numerical
simulations in Fig. \ref{tanner_final}. The FER curve changes the
slope, for higher $\alpha,$ where FER contribution from the
expression (\ref{other}) is not negligible.\begin{figure}[h]
\centering
\includegraphics[width=2.3in]{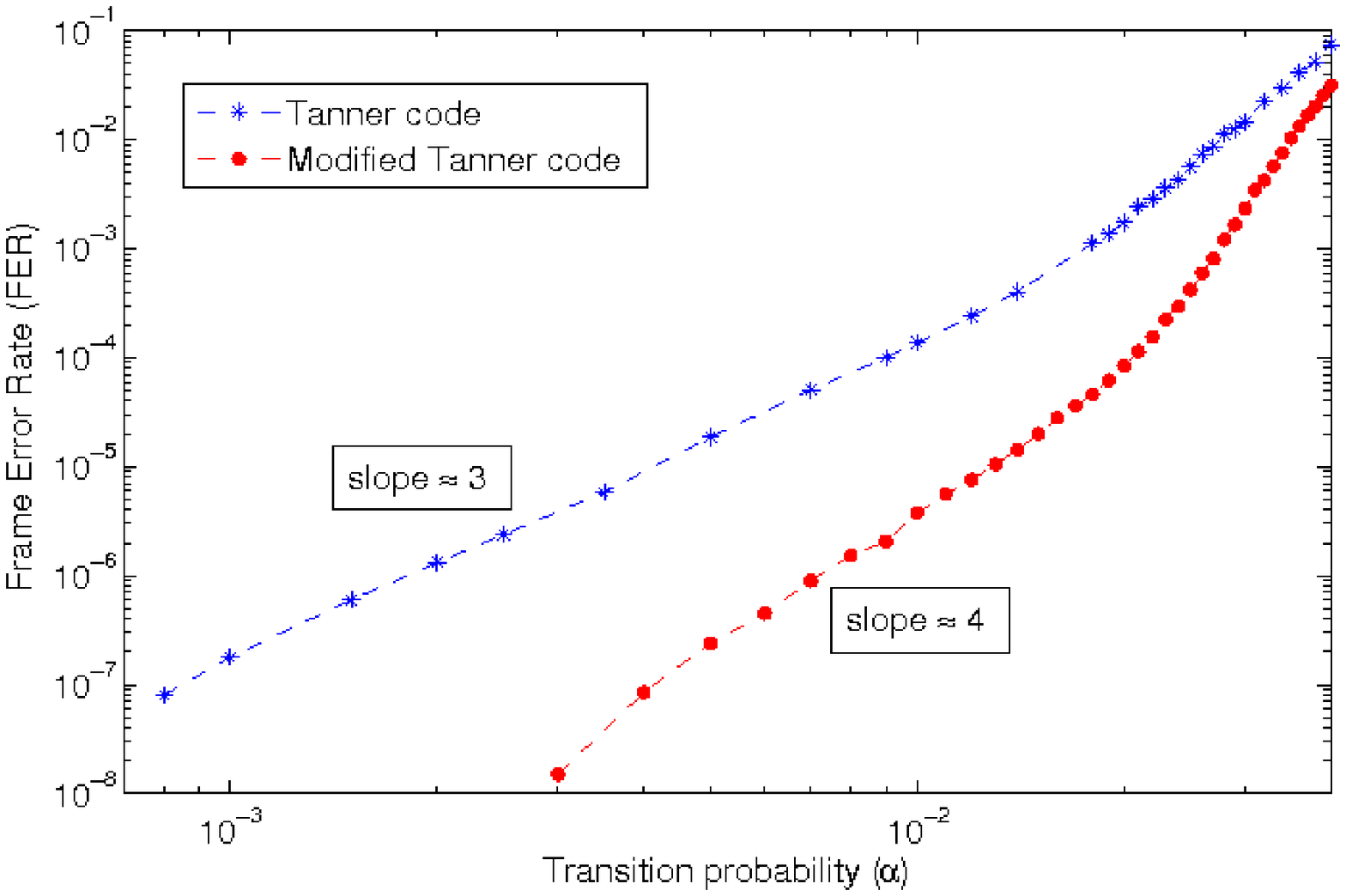}
\caption{Tanner code performance for a longer range of $\alpha$}
\label{tanner_final}
\end{figure} Note that there was a small  rate penalty to this procedure. The
original Tanner code has rate 0.4129, whereas the modified code has
rate 0.4065.  \end{ex}

\begin{ex} {(\it MacKay's (1008, 504) codes)} This is an example of how
the method can be used to produce better codes of a fixed length. We
have taken a 504 length MacKay  code and  constructed a  1008
($2*504$) length code. The new code performs better than MacKay
codes of length 1008.

Both original 504 and 1008 length codes have two types of trapping
sets with critical number $k=3,$ (5,3) and (3,3) (six cycles). We
ran the algorithm so that all (3,3) trapping sets are eliminated
from the newly constructed, but none of the (5,3) trapping sets. The
results are shown in Fig. \ref{MK}.\begin{figure}[h] \centering
\includegraphics[width=2.25in]{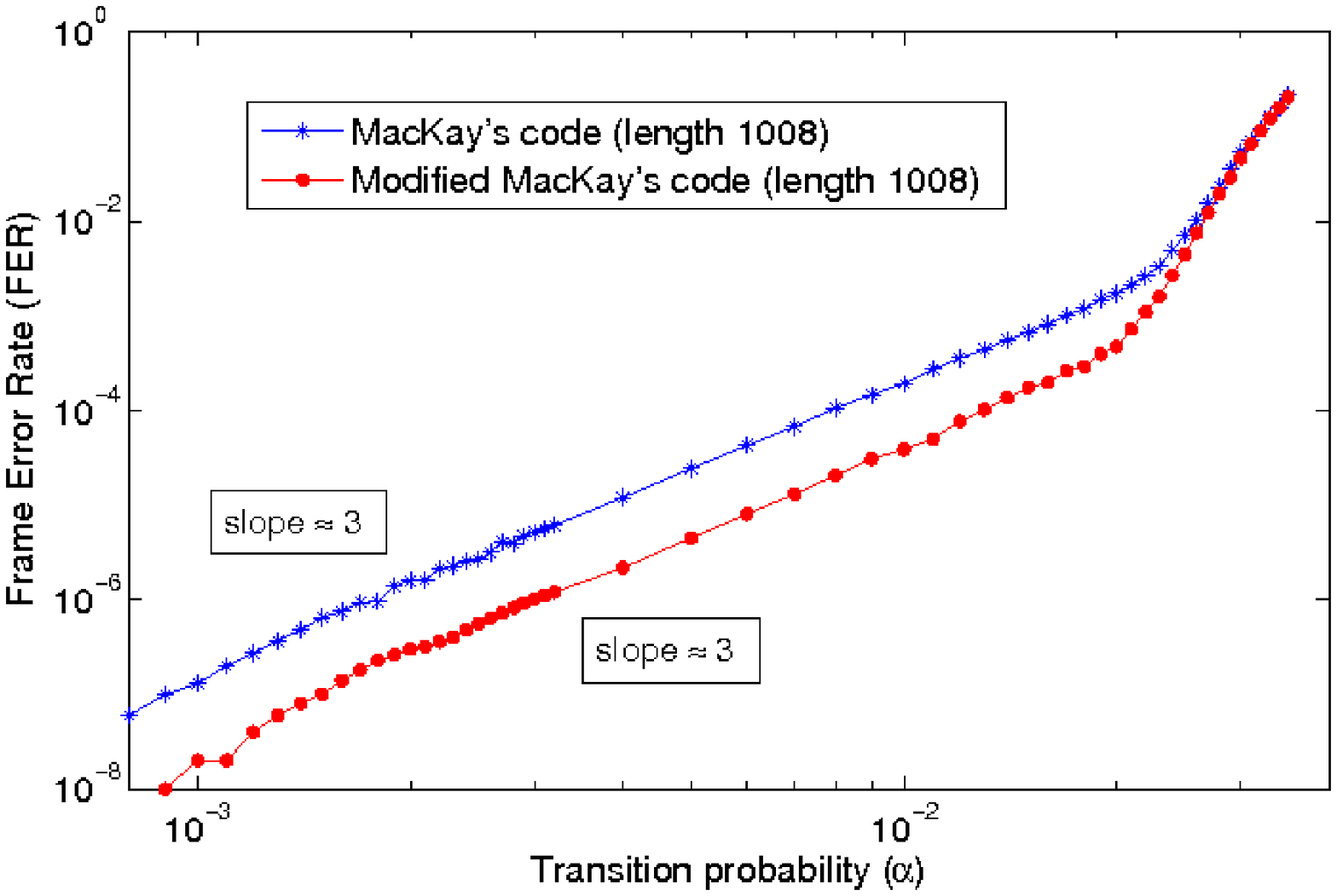}
\caption{MacKay's codes performance} \label{MK}
\end{figure}
It can be seen that, although the FER performance is improved, the slope of the FER curve is
approximately the same.\footnote{It is possible that a more sophisticated algorithm would also eliminate the (5,3) trapping sets. However, our goal with this example was to show  the performance when some, but not all, of the trapping sets with minimal critical number are eliminated.}
\end{ex}

\begin{ex}  {\it (AWGN channel)} This example illustrates two points. First is  that optimizing code for one decoding algorithm can lead to performance  improvement for other decoding algorithms. The  second point is that the use  of an appropriate axis scaling can greatly help in error floor analysis and code performance prediction.   

We present FER  results over AWGN channel and min-sum algorithm after 500 iterations for three codes, the original Tanner (155, 64) code, our modified Tanner (310, 126) from the Example 3  and a random (310, 127) code with column weight 3 and row weight 5. 

In the low SNR region,  where all kinds of error events are likely,  the length  (and rate) of a code govern the performance. In this region codes of length 310 have similar performance.   For high SNRs, however, code optimization in terms of trapping sets becomes important and random code performance becomes much worse than performance of the modified Tanner (310, 126) code. Notice a pronounced error floor for the random code. 

What is even more illustrative  is Fig. \ref{tanner} where we plot $\log(FER)$ versus SNR (not in dB) on the x-axis.  This is because for high SNRs on the AWGN channel, similarly to Eq. (\ref{other}),  $FER \propto \exp(-{\omega_{in} * \mbox{SNR}}/{2})$,  where $d_{in}$ is  pseudo-weight of the most likely error event. So on the graph with SNR on the x-axis which is not in dB, $\log(FER)$ curve will approach (from above) a straight line with slope equal to $-\omega_{in}/2$ as $\mbox{SNR}\rightarrow \infty$.  See \cite{wiberg} and  \cite{koetter} for further details.  Using these observations and numerical results obtained by simulations we can estimate that our modified code has  the slope approximately equal to 20, better than the original Tanner (155, 64) code with the slope of $\approx 14.$\footnote{The estimate for the Tanner code is  in accordance with  the pseudo-weight of the  single most likely error event  of $\approx 12.45$  reported in \cite{misha2}. }   

Further more, considering that the slope for the random code is $\approx 12$, we can claim that,  for SNR values higher than those on the plots, the Tanner code will perform better than the random code.

\begin{figure}[th]
\centering
\subfigure[ $\log(FER)$ versus SNR in dB] 
{   \label{dB} \includegraphics[width=2.2in]{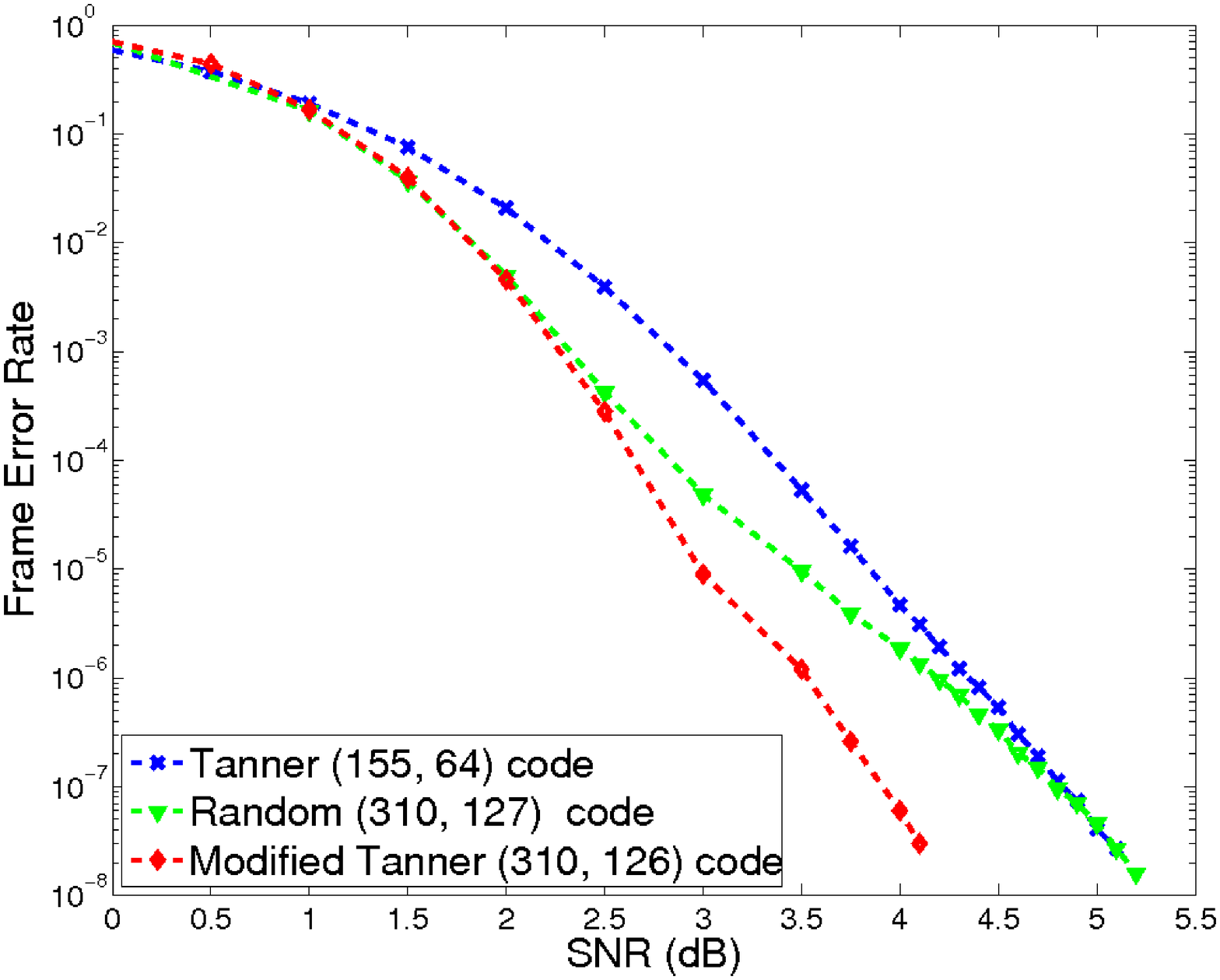} }
\subfigure[$\log(FER)$ versus SNR as $\frac{E}{N}$ (not in dB)]  
{ \label{tanner} \includegraphics[width=2.2in]{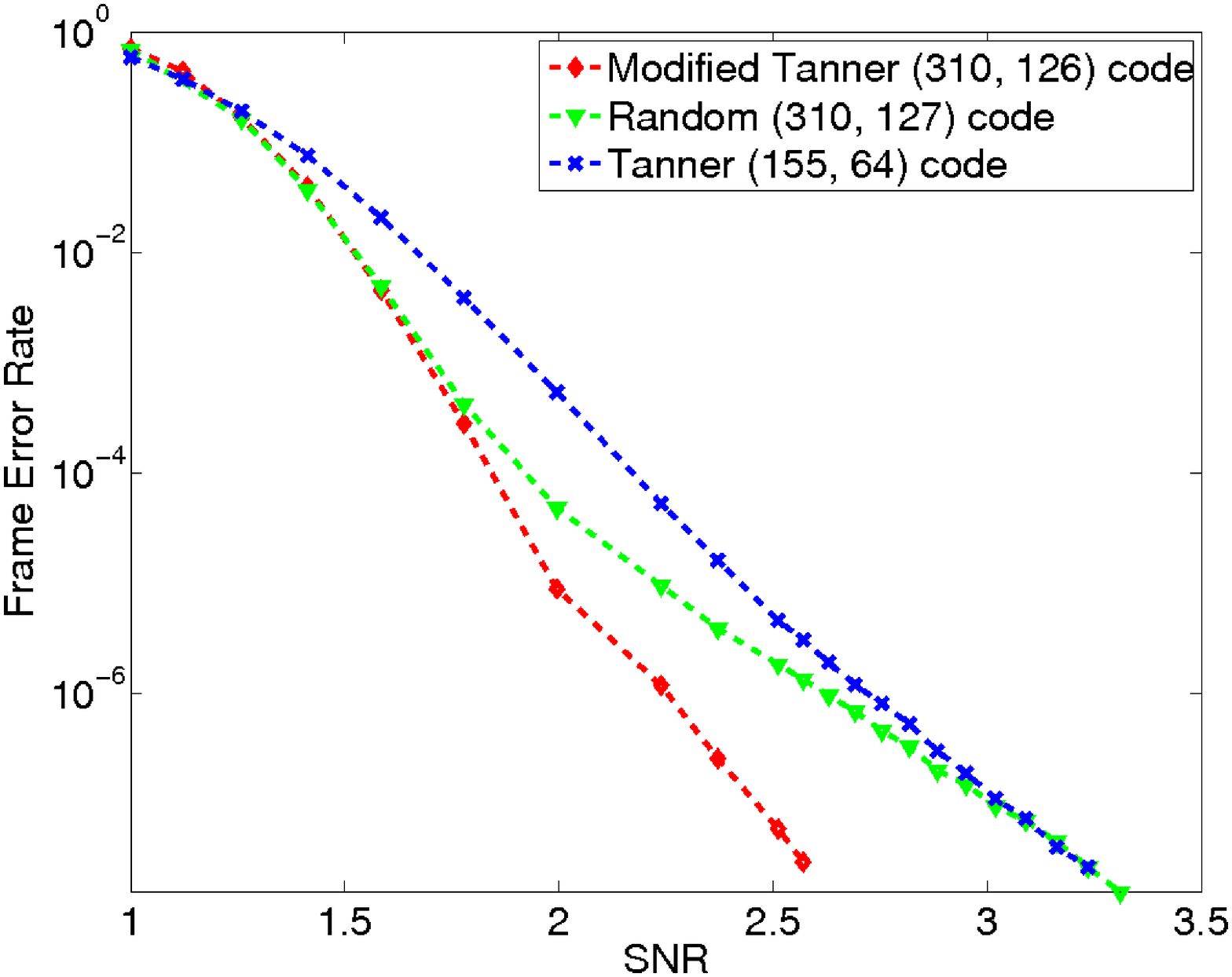} }
\caption{FER performance under min-sum decoding }
\end{figure}

\end{ex}

\section{Conclusion} \label{conclusion}

The proposed method allows the construction  of codes with good FER
performance, but low row/column weight (as opposed to projective
geometry codes) and therefore relatively low decoding complexity.
Although numerical results for the Gallager B decoder are presented, we
reiterate that the method can be used for code optimization based on
the trapping sets of an arbitrary decoder. 

The algorithm can   also be used to determine the pseudo-weight
spectrum of a code as follows. Once the most likely trapping sets
(those with the smallest pseudo-weight) are determined and
eliminated by the method,  the numerically obtained decoding performance
of a modified code, i.e., the slope of the FER curve with appropriate axis,  gives an estimate of the  pseudo-weight of the
next most likely trapping sets -just as it was done in the Example 5 with the Tanner code and the modified Tanner code.

\section*{Acknowledgment}
The authors would like to acknowledge valuable discussions with
Robert Indik, Misha Stepanov and Clifton Williamson.

\end{document}